\documentclass[11pt,a4paper]{article} 
\usepackage[symbol*]{footmisc}
\usepackage{array}
\pdfoutput=1 
\usepackage{amsmath}
\usepackage{amssymb}
\usepackage{amsfonts}
\usepackage{graphicx}

\def\bk{{\bf k}}
\def\bq{{\bf q}}
\def\la{~\mbox{\raisebox{-.6ex}{$\stackrel{<}{\sim}$}}~}
\def\ga{~\mbox{\raisebox{-.6ex}{$\stackrel{>}{\sim}$}}~}

\begin{document}

\thispagestyle{empty}
\vspace*{1cm}
\begin{center}
{\Large \bf {Parity violation in the Cosmic Microwave Background}}\\ 
\vspace*{0.3cm}
{\Large \bf {from a pseudoscalar inflaton}}\\
\vspace*{1.5cm} {\large Lorenzo Sorbo\footnote{\tt sorbo@physics.umass.edu}}\\
\vspace{.5cm}  
{\em Department of Physics, University of Massachusetts, Amherst, MA 01003}\\
\vspace{.15cm}
\vspace{1.5cm} 

ABSTRACT

\end{center}

{
If the inflaton $\phi$ is a pseudoscalar, then it naturally interacts with gauge fields through the coupling $\propto\phi\,F_{\mu\nu}\,\tilde{F}^{\mu\nu}$. Through this coupling, the rolling inflaton produces quanta of the gauge field, that in their turn source the tensor components of the metric perturbations. Due to the parity-violating nature of the system, the right- and the left-handed tensor modes have different amplitudes. Such an asymmetry manifests itself in the form of non-vanishing TB and EB correlation functions in the Cosmic Microwave Background (CMB). We compute the amplitude of the parity-violating tensor modes and we discuss two scenarios, consistent with the current data, where parity-violating CMB correlation functions will be detectable in future experiments.}

\vskip2.5cm

\begin{flushleft}
\end{flushleft}

\vfill \setcounter{page}{0} \setcounter{footnote}{0}

\section{Introduction}

Despite decades of model building, we are still looking for a preferred, compelling, UV-complete model of inflation. Some scenarios, however, appear to be more promising than others. In particular, due to their radiative stability, models where the inflaton is a pseudo-Nambu-Goldstone boson (pNGb~\cite{Freese:1990rb}) have received a significant amount of interest (see e.g.~\cite{models}). PNGbs are pseudoscalar particles: as they roll down their potential, they provide a macroscopic source of parity violation. The goal of this paper is to present a simple mechanism able to imprint such parity violation on the Cosmic Microwave Background.

A pseudoscalar inflaton $\phi$ (especially a pNGb, that is characterized by a broken shift symmetry) is generically expected to interact with gauge fields through the coupling $\phi\,F_{\mu\nu}\,\tilde{F}^{\mu\nu}/f$. As $\phi$ rolls down its potential, it provides a time-dependent background  for the quantization of the gauge field, amplifying its vacuum fluctuations into classical modes~\cite{Garretson:1992vt}, that in their turn are a source of gravitational waves. Because of the parity-violating nature of the system, only photons of a given helicity are produced~\cite{Anber:2006xt}, implying that gravitational waves of different helicities are produced with different amplitude. This way, parity violation in the inflaton sector manifests itself as a different amplitude of the power spectrum of the left-handed modes ${\cal P}^{t,-}$ of the graviton with respect to that, ${\cal P}^{t,+}$, of the right-handed ones. A measure of the net handedness of the tensor modes is the parameter $\Delta\chi\equiv \left({\cal P}^{t,+}-{\cal P}^{t,-}\right)/\left({\cal P}^{t,+}+{\cal P}^{t,-}\right)$~\cite{Saito:2007kt,Gluscevic:2010vv}. As we will see, in this scenario the value of $\Delta\chi$ will depend only on the value of the Hubble parameter $H$ during inflation and (exponentially) on the combination $\dot\phi/(H\,f)$, and can easily attain values indistinguishable from unity.

How do parity-violating tensor modes show up in the CMB? Tensor modes produced during inflation leave a trace on the CMB in the form of B-modes, the divergence-free component of the polarized radiation. While  B-modes have not been yet detected, the sensitivity of CMB experiments to these modes will improve substantially in the next years. B-modes are parity-odd, whereas E-modes (the curl-free component of the polarized CMB radiation) as well as the temperature fluctuations are parity-even. As a consequence, a nonvanishing  $\langle B\,E\rangle$  or $\langle B\,T\rangle$ correlation will signal parity violation in the CMB~\cite{Lue:1998mq}. In the best case scenario, values of $\Delta\chi$ as small as $.3$ might be detected at the $3\,\sigma$ level in a cosmic-variance limited experiment~\cite{Saito:2007kt,Gluscevic:2010vv}. 

Of course, in order to be viable, the model has to satisfy all experimental constraints. In this system, the main constraint comes from the requirement that nongaussianities are below the observational limits~\cite{Barnaby:2010vf}, and originates from the fact that the source of chiral tensor modes is also a source of scalar modes. Since these modes arise from a second order effect, they are intrinsically nongaussian. We discuss two scenarios where this constraint does not apply. The first option is that most of the density perturbations are generated by a second field (a curvaton~\cite{curvaton}) with gaussian perturbations. A second possibility is that the system contains several gauge fields. In both cases nongaussianities can be reduced to a safely small value while the tensor modes maintain an amplitude large enough to allow detectability.

The possibility of parity violation in the CMB was already considered in the past (see e.g.~\cite{Lue:1998mq,chiralgrav}). The mechanisms analyzed in those papers, however,  are based on the assumption of a parity violating term in the gravitational sector, whereas the present work relies only on parity violation in the matter sector of the theory. Note also that the magnitude of parity violation induced by a gravitational Chern Simons term is limited to small values, according to~\cite{cherneft}, once one requires validity of the field theoretical description of the system. 

The paper is organized as follows. Section 2 contains a review of the production of helical gauge modes by an axion-like inflaton. The generation of parity-violating gravitational waves is discussed in section 3. Section 4 deals with the constraints on the parameter space and describes two scenarios, consistent with the current constraints, where parity violating correlation functions would be detectable in future CMB surveys. We conclude in section 5.

\section{Production of helical gauge fields by a pseudoscalar inflaton}%

In order to make this paper self-contained, this section reviews the equations describing the production of helical modes of a $U(1)$ gauge field coupled to a pseudoscalar inflaton $\phi$. A more detailed presentation of the results presented in this section can be found in~\cite{Anber:2006xt}. The Lagrangian density of our system is given by
\begin{equation}\label{lag}
{\cal L}=-\frac{1}{2}\left(\partial\phi \right)^2-V(\phi)-\frac{1}{4}F_{\mu\nu}F^{\mu\nu}-\frac{\phi}{4\,f}\, F_{\mu\nu}\tilde F^{\mu\nu},
\end{equation}
where $V(\phi)$ is an arbitrary potential able to support slow-roll inflation. The dimensionful parameter $f$ is a measure of the coupling of $\phi$ to the gauge field.

In terms of the vector potential ${\bf A}\left(\tau,\,{\bf x}\right)$, defined by $a^2\,{\bf B}=\nabla\times{\bf A}$, $a^2\,{\bf E}=-{\bf A}'$, and neglecting the spatial gradients of $\phi$, the equations for the gauge field read
\begin{eqnarray}
\label{a15}
\left(\frac{\partial^{2}}{\partial \tau^{2}}-\nabla^{2}-\frac{\phi'}{f}\,\nabla\times \right){\bf A}=0,\,\,\, \qquad\nabla\cdot{\bf A}=0\,,
\end{eqnarray} 
where the prime denotes differentiation with respect to the conformal time $\tau$ and $a(\tau)$ is the scale factor of the flat Friedmann-Robertson-Walker Universe .

In order to study the generation of the electromagnetic field induced by the rolling pseudoscalar, we promote the classical field ${\bf A}(\tau,\,{\bf x})$ to an operator $\hat {\bf A}\left(\tau,\,{\bf x}\right)$ that we decompose into annihilation and creation operators $\hat{a}_\lambda^\bk$, $\hat{a}_\lambda^\bk{}^\dagger$
\begin{eqnarray}\label{a16}
{\hat A}_i(\tau,\,{\bf x})=\int\frac{d^3{\bf k}}{\left(2\pi \right)^{3/2}}e^{i{\bf k\cdot x}}{\hat A}_i(\tau,{\bf k})=\sum_{\lambda=\pm}\int \frac{d^3\bk}{\left(2\pi \right)^{3/2}}\left[\epsilon^i_\lambda(\bk)\,A_\lambda(\tau,\,\bk)\,{\hat a}_\lambda^{\bk}\,e^{i{\bf k\cdot x}}+{\mathrm {h.c.}}\right],
\end{eqnarray}
where the helicity vectors $\epsilon^i_\pm$ are defined so that $k_i\,  \epsilon^i_\pm=0$, $\varepsilon_{abc}\,k_b\,\epsilon^c_\pm=\mp i\,k\,\epsilon^c_\pm$, $\epsilon^i_\pm\,\epsilon^i_\mp=1$ and $\epsilon^i_\pm\,\epsilon^i_\pm=0$. Then, the functions $A_\pm$ must satisfy the equation $A_{\pm}''+(k^2 \mp k\,\phi'/f)A_{\pm}=0$.

Since we are working on an inflating background, we assume de Sitter metric $a\left(\tau\right)\simeq -1/(H\,\tau)$, and $\phi'/a=\sqrt{2\,\epsilon}\,H\,M_P\simeq\,$constant. Hence, the equation for $A_\pm$ reads
\begin{equation}
\label{d3}
\frac{d^{2}A_\pm(\tau,\, k)}{d\tau^{2}}+\left[k^{2}\pm 2\,k\,\frac{\xi}{\tau} \right]A_\pm(\tau,\, k)=0\mbox{ ,}
\end{equation} 
where we have defined 
\begin{equation}
\xi\equiv\frac{\dot\phi}{2\,f\,H}=\sqrt{\frac{\epsilon}{2}}\,\frac{M_P}{f}\,\,,
\end{equation}
We will be interested in the case $\xi\ga{\cal {O}}\left(1\right)$.

Depending on the sign of $\xi$, one of the two modes $A_{+}$ or $A_{-}$ in (\ref{d3}) develops an instability (we assume without loss of generality that $\xi>0$). The other mode stays essentially in vacuum. The difference between the amplitude of the left- and that of the right-handed photons shows that the gauge modes have inherited the parity violating nature of the rolling inflaton.

The solution of equation~(\ref{d3}) that reduces to positive frequency in the limit $k\,\tau\rightarrow -\infty$ is $A_\pm(\tau,\, k)=\frac{1}{\sqrt{2\,k}}[i\,F_0(\pm\xi,\,- k\,\tau)+G_0(\pm\xi,\,- k\,\tau)]$, where $F_0$ and $G_0$ are the regular and irregular Coulomb wave functions. The positive-helicity mode is rapidly amplified, and $A_+$ peaks at momenta $k$ for which $\left(8\,\xi\right)^{-1}\la |k\,\tau|\ll 2\,\xi$, where it is well approximated by
\begin{equation}
\label{approx1}
A_+(\tau,\, k)\simeq 
\frac{1}{\sqrt{2\,k}}\left(\frac{ k}{2\,\xi\,aH}\right)^{1/4}e^{\pi\,\xi-2\,\sqrt{2\xi \,k/aH}}\,.
\end{equation}

$A_+$ is thus amplified by a factor $e^{\pi\,\xi}$. On the other hand, the modes $A_-$ are not amplified by the rolling inflaton, and from now on we ignore them.

\section{Generation of tensor modes by the gauge field}%

In this section we study the production of gravitational waves induced by the electromagnetic modes. Since we are focusing only on the tensor modes, our Ansatz for the metric is
\begin{equation}
ds^2=a^2(\tau)\,\left[-d\tau^2+\left(\delta_{ij}+h_{ij}\right)\,dx^i\,dx^j\right],
\end{equation}
with $h_i{}^i=h_{ij},_j=0$. The equation of motion for $h_{ij}$ reads
\begin{equation}\label{eqh}
h_{ij}''+2\,\frac{a'}{a}\,h_{ij}'-\Delta\,h_{ij}=\frac{2}{M_P^2}\,\Pi_{ij}{}^{lm}\,T^{EM}_{lm}
\end{equation}
where $\Pi_{ij}{}^{lm}=\Pi^i_l\,\Pi^j_m-\frac{1}{2}\Pi_{ij}\,\Pi^{lm}$ is the transverse traceless projector, with $\Pi_{ij}=\delta_{ij}-\partial_i\,\partial_j/\Delta$ and where $T^{EM}_{lm}$ represents the spatial part of the stress-energy tensor of the gauge field $T^{EM}_{ij}=-a^2\left(E_i\,E_j+B_i\,B_j\right)+(\dots)\,\delta_{ij}$. Next, we go to momentum space, project $h_{ij}$ on positive- and negative-helicity modes as
\begin{equation}
h^{ij}(\bk)=\sqrt{2}\,\sum_{\lambda=\pm}\epsilon_\lambda^i(\bk)\,\epsilon_\lambda^j(\bk)\,h_\lambda(\tau,\,\bk)\,\,,
\end{equation}
and introduce the polarization tensors $\Pi_\pm^{ij}({\bk})=\epsilon_\mp^i({\bk})\,\epsilon_\mp^j({\bk})/\sqrt{2}$, so that $h_\pm({\bk})=\Pi^{ij}_\pm({\bk})\,h_{ij}({\bk})$. We can now promote the functions $h_\pm$ to operators $\hat{h}_\pm$. Using $\Pi_\pm^{ij}\,\Pi_{ij}{}^{lm}=\Pi_\pm^{lm}$, and neglecting for the time being the solution of the homogeneous part of eq.~(\ref{eqh}), the expression of ${\hat h}_\pm$ can be found using the techniques of~\cite{Barnaby:2009mc}
\begin{eqnarray}
&&{\hat h}_\pm(\bk)=-\frac{2\,H^2}{M_P^2}\int d\tau'\,G_k(\tau,\,\tau')\,\tau'{}^2\int\frac{d^3{\bq}}{(2\pi)^{3/2}}\,\Pi_\pm^{lm}({\bk})\times\nonumber\\
&&\times\left[\hat{A}_l'({\bq},\tau')\,\hat{A}_m'({\bk}-{\bq},\tau')-\epsilon_{lab}\,q_a\,\hat{A}_b({\bq},\tau')\,\varepsilon_{mcd}\,(k_c-q_c)\,\hat{A}_d({\bk}-{\bq},\tau')\,\right].
\end{eqnarray}
where we have defined the retarded Green function for the operator $d^2/d\tau^2-(2/\tau)d/d\tau+k^2$ 
\begin{equation}
G_k(\tau,\tau')=\frac{1}{k^3\,\tau'{}^2}\left[\left(1+k^2\,\tau\,\tau'\right)\sin k\left(\tau-\tau'\right)+k\left(\tau'-\tau\right)\,\cos k\left(\tau-\tau'\right)\right]\,, 
\end{equation}
for $\tau>\tau'$, while $G_k(\tau<\tau')=0$

Next, we set $A_-=0$ and, since most of the production of tensor modes happens for $\left(8\,\xi\right)^{-1}\ll |k\tau|\ll 2\,\xi$, we use the approximation~(\ref{approx1}). By applying Wick's theorem, we find that, for $\xi\ga 1$, the two point function of the helicity-$\lambda$ graviton is given by
\begin{eqnarray}\label{longint}
&&\langle h_\lambda({\bk})\,h_\lambda({\bk}')\rangle=\frac{H^4\,\xi}{4\,\pi^3\,M_P^4}e^{4\pi\xi}\,\delta(\bk+\bk')\,\int d\tau'\,d\tau''\,|\tau'|^{3/2}\,|\tau''|^{3/2}G_k(\tau,\,\tau')\,G_k(\tau,\,\tau'')\,\times\\
&&\times\int d^3\bq\left|\epsilon^i_{-\lambda}({\bk})\,\epsilon^i_+({\bq})\right|^2\left|\epsilon^j_{-\lambda}({\bk})\,\epsilon^j_+(\bk-\bq)\right|^2\,{\sqrt{|\bk-\bq|}}\,{\sqrt{q}}\,e^{-2\sqrt{2\,\xi}(\sqrt{|\tau'|}+\sqrt{|\tau''|})\,\left(\sqrt{q}+\sqrt{|\bk-\bq|}\right)}\,\nonumber,
\end{eqnarray}
where the first line depends on the propagators while the second line depends on the amplitude of the gauge field and on the helicity of the graviton. The second line can be written more explicitly after using the property of the helicity projectors
\begin{equation}
\left|\epsilon^i_{-\lambda}({\bf p}_1)\,\epsilon^i_+({\bf p}_2)\right|^2=\frac{1}{4}\left(1+\lambda\,\frac{{\bf p}_1\cdot {\bf p}_2}{p_1\,p_2}\right)^2\,.
\end{equation}

In the large scale limit $-k\,\tau\rightarrow 0$ the integral~(\ref{longint}) can be computed numerically.  Rather than plotting the result of the numerical integration, we give the following analytical approximation that for $\xi\gtrsim 3$ (that, as we will see, is the regime we are interested in) is good at the $15\%$, and rapidly improves as $\xi$ increases
\begin{eqnarray}
&&\langle h_+({\bk})\,h_+({\bk}')\rangle\simeq 8.6\times 10^{-7}\,\frac{H^4}{M_P^4}\frac{e^{4\pi\xi}}{\xi^6}\,\frac{\delta(\bk+\bk')}{k^3},\nonumber\\
&&\langle h_-({\bk})\,h_-({\bk}')\rangle\simeq 1.8
\times 10^{-9}\,\frac{H^4}{M_P^4}\frac{e^{4\pi\xi}}{\xi^6}\,\frac{\delta(\bk+\bk')}{k^3}.
\end{eqnarray}

We thus see that the spectra of both the left- and the right-handed tensor modes are scale invariant. As a consequence of the violation of parity, however, their amplitude differs by a factor $\sim 10^3$. The numerical discrepancy between the two spectra originates from  the term $\left|\epsilon^i_{-\lambda}({\bk})\,\epsilon^i_+({\bq})\right|^2\left|\epsilon^j_{-\lambda}({\bk})\,\epsilon^j_+(\bk-\bq)\right|^2$ in eq.~(\ref{longint}). In particular, the relations $\epsilon^i_\pm\,\epsilon^i_\mp=1$ and $\epsilon^i_\pm\,\epsilon^i_\pm=0$ imply that in the limit $|\bq|\ll|\bk|$ this term vanishes if $\lambda=-$, but stays finite for $\lambda=+$. Physically, this can understood by observing that for a small transverse momentum, conservation of angular momentum does not allow two positive-helicity photons to generate a negative-helicity graviton.

Of course, one should also take into account the parity-symmetric component of gravitons that are generated by the usual amplification of vacuum fluctuations in de Sitter space and that correspond to the solutions of the homogeneous part of eq.~(\ref{eqh}). These are uncorrelated with those discussed above, so that the overall left- and right-handed power spectra read
\begin{eqnarray}\label{main}
&&{\cal {P}}^{t,+}=\frac{H^2}{\pi^2\,M_P^2}\,\left(1+8.6\times 10^{-7}\,\frac{H^2}{M_P^2}\,\frac{e^{4\,\pi\,\xi}}{\xi^6}\right)\,\,,\nonumber\\
&&{\cal {P}}^{t,-}=\frac{H^2}{\pi^2\,M_P^2}\,\left(1+1.8\times 10^{-9}\frac{H^2}{M_P^2}\,\frac{e^{4\,\pi\,\xi}}{\xi^6}\right)\,\,,
\end{eqnarray}
from which we extract the chirality parameter
\begin{equation}\label{deltachi}
\Delta\chi=\frac{4.3\times 10^{-7}\,\frac{e^{4\,\pi\,\xi}}{\xi^6}\,\frac{H^2}{M_P^2}}{1+4.3\times 10^{-7}\,\frac{e^{4\,\pi\,\xi}}{\xi^6}\,\frac{H^2}{M_P^2}}\,.
\end{equation}

\section{Discussion}

Let us now discuss the constraints on the model along with the prospects of observing such a chiral background of gravitational waves. Our main result~(\ref{main}) depends only on the two parameter $H$ and $\xi$. An extra parameter, the slow roll parameter $\epsilon$, appears when we study the observational constraints on our scenario. Therefore, the entire system is in principle described by a three-dimensional parameter space. It is possible to eliminate one parameter by imposing COBE normalization of the spectrum of scalar perturbations. The amplitude of the scalar perturbations is also affected by the presence of the excited electromagnetic modes, and has been computed for this system in~\cite{Barnaby:2010vf}, that has obtained the following expression (accurate at the $25\%$ level for $\xi\gtrsim 3$)
\begin{equation}\label{pzeta}
{\cal {P}}^\phi_\zeta=\frac{H^2}{8\,\pi^2\,\epsilon\,M_P^2}\,\left[1+9.5\times 10^{-7}\,\frac{H^2}{\epsilon\,M_P^2}\,\frac{e^{4\,\pi\,\xi}}{\xi^6}\right]\,,
\end{equation}
where COBE normalization imposes the condition ${\cal {P}}^\phi_\zeta={\cal {P}}^{\mathrm {obs}}_\zeta=2.5\,\times 10^{-9}$. 

The strongest constraint on the system comes from the requirement that nongaussianities are within the limits set by observations. Nongaussianities in this setting have also been studied by~\cite{Barnaby:2010vf}, where it was found that the bispectrum has maximal amplitude in the case of equilateral configurations, where 
\begin{equation}
f_{NL}^{\mathrm {equil}}\simeq 8.9\,\times 10^{4}\,\frac{H^6}{\epsilon^3\,M_P^6}\frac{e^{6\,\pi\,\xi}}{\xi^9}\,.
\end{equation}
From the current WMAP limit~\cite{Komatsu:2010fb} $f_{NL}^{\mathrm {equil}}<266$ we derive the bound $\xi<2.6$. For such small values of $\xi$, and for $H\la 10^{-4}\,M_P$, the quantity $\Delta\chi$ is tiny and unobservable. We thus conclude that the simplest version of this scenario is unable to yield detectable parity violation while respecting the constraints from nongaussianities. Let us now discuss two modifications of this scenario where parity violation {\em can} be detected while complying with the other observations.

\subsection{A curvaton}

A first possibility is to assume that most of the density perturbations is provided by a second scalar field, a curvaton~\cite{curvaton}, that obeys an essentially  gaussian statistics. In this case, equation~(\ref{pzeta}) gives only the contribution of the inflaton $\phi$ to the observed power spectrum, that we require to be smaller than the observed value ${\cal {P}}_\zeta^{\mathrm {obs}}=2.5\,\times 10^{-9}$. 

In order to get a feel of the behavior of the system in this case, let us first consider the regime of large $\xi$, where the first term in brackets in eqs.~(\ref{main}) and~(\ref{pzeta}) is negligible with respect to the second term, that has an exponential dependence on $\xi$.

Let us then denote by $\delta<1$ the fraction of contribution from eq.~(\ref{pzeta}) to the observed ${\cal {P}}_\zeta^{\mathrm {obs}}$, so that ${\cal {P}}_\zeta^\phi= 2.5\,\delta\times10^{-9}$. It is easy to see that in this case $f_{NL}^{\mathrm {equil}}\simeq 8400\,\delta^{3/2}$,  implying that $\delta\simeq 0.1$ is already sufficient to make $f_{NL}$ compatible with observations for all values of $\xi$. As a consequence of the presence of a curvaton, also the tensor-to-scalar ratio is reduced by a factor of $\delta$. Since we are considering the limit $\xi\gg 1$, the tensor modes are fully chiral ($\Delta\chi\simeq 1$), and the tensor-to-scalar ratio $r=\left({\cal P}^{t,-}+{\cal P}^{t,+}\right))/{\cal {P}}_\zeta^{\mathrm {obs}}$ evaluates to $r\simeq 7.2\times\delta\times\epsilon^2$, where $\epsilon=M_P^2\,V,_\phi^2/2\,V^2$ is the slow-roll parameter during inflation and is unrelated to the parameters in the curvaton sector. For a fully chiral spectrum of gravitational waves, $r$ as small as $\sim 0.009$ might lead to a detectable parity violation, at 95\% confidence level, in a cosmic-variance-limited CMB experiment~\cite{Gluscevic:2010vv}. This corresponds, for $\delta=0.1$ to requiring $\epsilon\ga 1/9$. While this would be a rather large value of the slow-roll parameter in the usual models of inflation, it is worth recalling that in curvaton models the only requirement is that $\epsilon\la 1$ for inflation to occur. In fact, the constraint on the spectral index $n=0.963\pm 0.012$~\cite{Komatsu:2010fb} can be always satisfied by $n-1=-2\,\epsilon+2\,\eta_{C}$, where $\eta_C$, that is related to the second derivative of the curvaton potential~\cite{curvaton}, can be tuned to compensate for the contribution to $n-1$ by $\epsilon$.

In order to clarify the behavior of the system for all values of $\xi$, we plot in figure 1 the parameter space of the model for the choice $\delta=0.1$. The region below the thicker line leads to a violation of parity too small to be detected. The region above the thinner line corresponds to a value of $r$ larger than $0.24$, excluded by~\cite{Komatsu:2010fb}. Figure 1 shows that for $\xi\ga 3$ there is a region in parameter space consistent with the current data that would lead to a detection of parity violation in the CMB in the future surveys.

\subsection{Many gauge fields}

It is also possible to reduce the level of nongaussianities by considering a system with several gauge fields. Let us denote by ${\cal {N}}$ the number of $U(1)$ gauge fields coupled to the inflaton, all with the same coupling $1/f$, in the Lagrangian~(\ref{lag}). Eqs.~(\ref{main}),~(\ref{deltachi}) and~(\ref{pzeta}) are then modified by multiplying by ${\cal {N}}$ their $\xi$-dependent parts. Also the expression giving $f_{NL}^{\mathrm {equil}}$ must be modified by multiplying it by a factor of ${\cal{N}}$. Since, in the regime of large $\xi$, both the two- and the three-point functions are proportional to ${\cal {N}}$, once COBE normalization is imposed the three point function scales as $1/\sqrt{{\cal {N}}}$.  As a consequence, by setting ${\cal {N}}\ga 10^3$, the constraint from nongaussianities is satisfied for all values of $\xi$. While a system with $\sim 10^3$ gauge fields does not appear to be very natural, it is not even unreasonable -- it is indeed rather common to run into string compactifications with thousands of degrees of freedom. Just to mention one possibility, the ${\cal {O}}(10^3)$ $A_{\mu}$s could correspond to the modes of a weakly coupled nonabelian gauge group emerging from a stack of $\sim 30$ branes. If the gauge sector is large enough to nullify the nongaussianity constraint, then the parameter space of the system looks very similar to the one depicted in figure 1. In this case, however, one finds that in the regime of large $\xi$, $r$ does not pay the suppression by a factor of $\delta$ found in the previous subsection. As a consequence, a value of $\epsilon$ as small as $1/30$ can lead to a spectrum of chiral gravitational waves that might be detected at 95\% confidence level in a cosmic-variance limited experiment.

\begin{figure*}[t]
\centering
    \includegraphics[width=.7\textwidth]{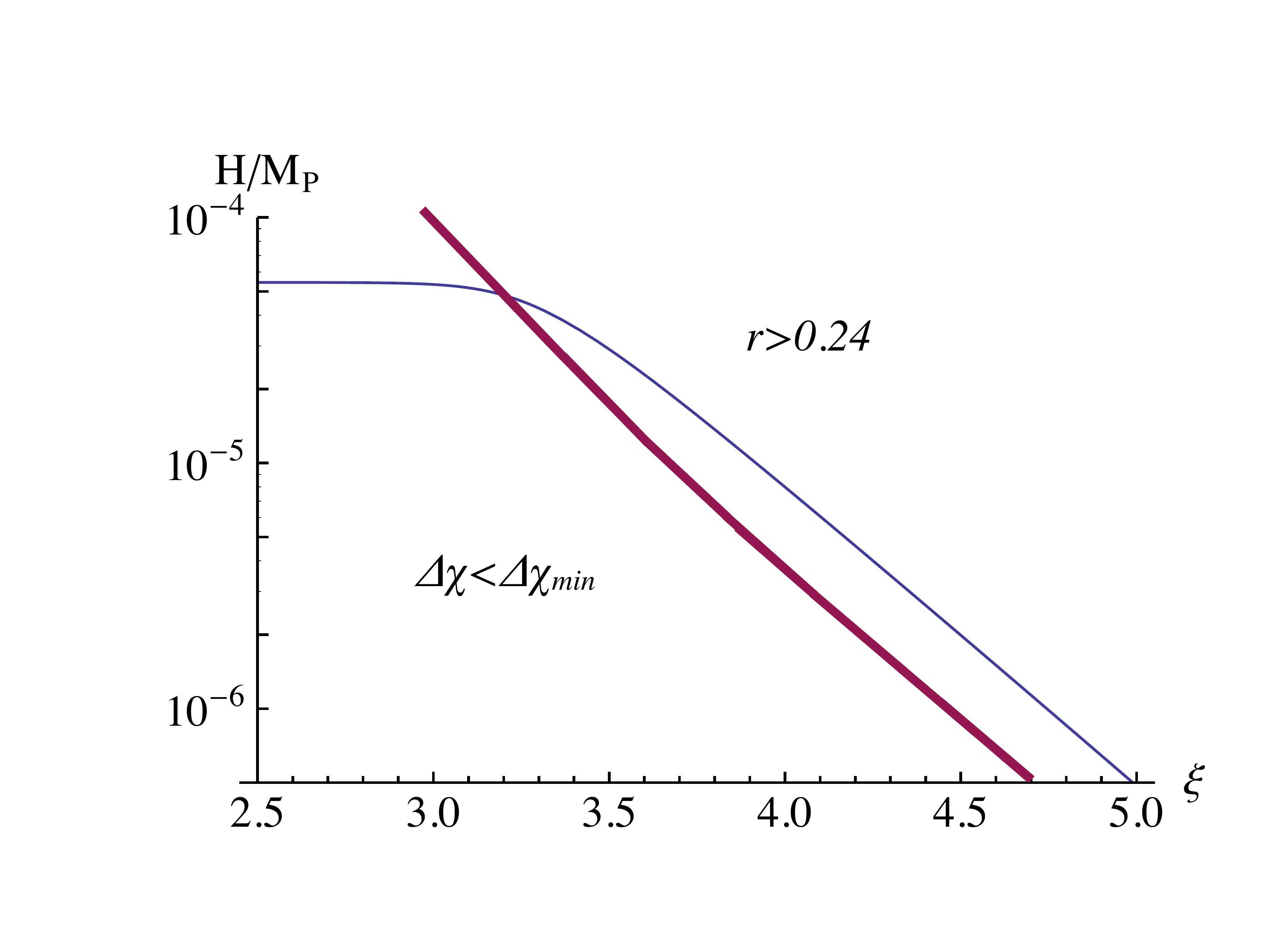}
    \caption{Parameter space for parity violating tensor modes. The choice of parameters in the plot corresponds to the case where 90\% of the scalar power spectrum is provided by a curvaton (i.e. $\delta=0.1$, see section 4.1 for details). In the region to the left of the thicker line, marked by $\Delta\chi<\Delta\chi_{min}$, the effect is too small to be observed at the 95\% confidence level in a cosmic-variance limited experiments. In the region to the right of the thinner line, marked by $r>0.24$, the amplitude of the tensor modes exceeds the current limits. The region between the two lines and below $H\simeq 5\times 10^{-5}\,M_P$ is allowed by current data and leads to observable parity violation in the CMB.}
\end{figure*}

\section{Conclusions}%

The main result of this paper is given by eqs.~(\ref{main}), that show that a pseudoscalar inflaton, through its natural coupling to gauge fields, can induce a parity-violating component in the spectrum of gravitational waves.  The degree of chirality~(\ref{deltachi}) depends exponentially on the quantity $\sqrt{\epsilon}\,M_P/f$, implying the spectrum of primordial gravitational waves is chiral in a large portion of the parameter space. 

This scenario represents, to our knowledge, the first example where parity violation can be imprinted on the CMB without invoking new physics in the gravitational sector.

Since the mechanism responsible for the generation of chiral tensor modes does also generate an intrinsically non-gaussian component of the scalar perturbations, the simplest version of this scenario cannot produce observable parity violation without exceeding the bounds imposed by non-observation of nongaussianities~\cite{Barnaby:2010vf}. In section 4, we have however presented two systems where such bounds do not apply, and that provide a proof of the existence of consistent models, compatible with current observations, that might lead to observable violation of parity in the CMB.

A few additional comments are in order. In this scenario there is no simple proportionality between the value of the tensor-to-scalar ratio $r$ and the Hubble parameter during inflation. In particular, one can see from figure 1 that, for $\xi\gtrsim 3$, $r$ could be observable even for very small values of the Hubble parameter. 

Also, it is worth noting that parity-violating correlation functions in the CMB could also emerge as an effect of birefringence due to a pseudoscalar quintessence field~\cite{Carroll:1998zi}. Ref.~\cite{Gluscevic:2010vv}, however, has shown that it would be possible to discriminate parity-violating correlation functions of primordial origin from those induced by dynamics after the last scattering. 

Finally, it would be interesting to study whether the system~(\ref{lag}) can generate a parity-odd component in the CMB bispectrum~\cite{Kamionkowski:2010rb}.

\smallskip

{\bf Acknowledgments.} It is a pleasure to thank John Donoghue, Nemanja Kaloper, Mikhail Voloshin and especially Marco Peloso for useful discussions. This work is partially supported by the U.S. National Science Foundation grant PHY-0555304.


\begin{thebibliography}{100}

\bibitem{Freese:1990rb}
  K.~Freese, J.~A.~Frieman and A.~V.~Olinto,
  Phys.\ Rev.\ Lett.\  {\bf 65}, 3233 (1990).

\bibitem{models}
 J.~E.~Kim, H.~P.~Nilles and M.~Peloso,
  JCAP {\bf 0501}, 005 (2005)
  [arXiv:hep-ph/0409138];
  S.~Dimopoulos, S.~Kachru, J.~McGreevy and J.~G.~Wacker,
  JCAP {\bf 0808}, 003 (2008)
  [arXiv:hep-th/0507205];
  L.~McAllister, E.~Silverstein and A.~Westphal,
  arXiv:0808.0706 [hep-th];
  N.~Kaloper and L.~Sorbo,
  Phys.\ Rev.\ Lett.\  {\bf 102}, 121301 (2009)
  [arXiv:0811.1989 [hep-th]];
  N.~Kaloper, A.~Lawrence and L.~Sorbo,
  arXiv:1101.0026 [hep-th];
  M.~M.~Anber and L.~Sorbo,
  Phys.\ Rev.\  D {\bf 81}, 043534 (2010)
  [arXiv:0908.4089 [hep-th]].

\bibitem{Garretson:1992vt}
  W.~D.~Garretson, G.~B.~Field and S.~M.~Carroll,
  Phys.\ Rev.\ D {\bf 46}, 5346 (1992)
  [arXiv:hep-ph/9209238].

\bibitem{Anber:2006xt}
  M.~M.~Anber and L.~Sorbo,
  JCAP {\bf 0610}, 018 (2006)
  [arXiv:astro-ph/0606534].

\bibitem{Saito:2007kt}
  S.~Saito, K.~Ichiki and A.~Taruya,
  JCAP {\bf 0709}, 002 (2007)
  [arXiv:0705.3701 [astro-ph]].

\bibitem{Gluscevic:2010vv}
  V.~Gluscevic and M.~Kamionkowski,
  Phys.\ Rev.\  D {\bf 81}, 123529 (2010)
  [arXiv:1002.1308 [astro-ph.CO]].

\bibitem{Lue:1998mq}
  A.~Lue, L.~M.~Wang and M.~Kamionkowski,
  Phys.\ Rev.\ Lett.\  {\bf 83}, 1506 (1999)
  [arXiv:astro-ph/9812088].

\bibitem{Barnaby:2010vf}
  N.~Barnaby and M.~Peloso,
  arXiv:1011.1500 [hep-ph].

\bibitem{curvaton}
  A.~D.~Linde and V.~F.~Mukhanov,
  Phys.\ Rev.\  D {\bf 56}, 535 (1997)
  [arXiv:astro-ph/9610219];
  D.~H.~Lyth and D.~Wands,
  Phys.\ Lett.\  B {\bf 524}, 5 (2002)
  [arXiv:hep-ph/0110002].

\bibitem{chiralgrav}
  S.~Alexander and J.~Martin,
  Phys.\ Rev.\  D {\bf 71}, 063526 (2005)
  [arXiv:hep-th/0410230];
  C.~R.~Contaldi, J.~Magueijo and L.~Smolin,
  Phys.\ Rev.\ Lett.\  {\bf 101}, 141101 (2008)
  [arXiv:0806.3082 [astro-ph]].
  
\bibitem{cherneft}
   D.~H.~Lyth, C.~Quimbay and Y.~Rodriguez,
  JHEP {\bf 0503}, 016 (2005)
  [arXiv:hep-th/0501153];
  M.~Satoh,
  JCAP {\bf 1011}, 024 (2010)
  [arXiv:1008.2724 [astro-ph.CO]].


\bibitem{Barnaby:2009mc}
  N.~Barnaby, Z.~Huang, L.~Kofman and D.~Pogosyan,
  arXiv:0902.0615 [hep-th].

\bibitem{Komatsu:2010fb}
  E.~Komatsu {\it et al.},
  arXiv:1001.4538 [astro-ph.CO].

\bibitem{Carroll:1998zi}
  S.~M.~Carroll,
  Phys.\ Rev.\ Lett.\  {\bf 81}, 3067 (1998)
  [arXiv:astro-ph/9806099];
  F.~Finelli, M.~Galaverni,
  Phys.\ Rev.\  {\bf D79}, 063002 (2009).
  [arXiv:0802.4210 [astro-ph]].

\bibitem{Kamionkowski:2010rb}
  M.~Kamionkowski and T.~Souradeep,
  arXiv:1010.4304 [astro-ph.CO].

\end{thebibliography}
\end{document}